\documentclass[twocolumn,preprintnumbers,amsmath,amssymb,showpacs,aps,superscriptaddress]{revtex4}
\usepackage{graphicx}
\usepackage{dcolumn}
\usepackage{bm}
\usepackage{amsmath}
\usepackage{amssymb}
\usepackage{epsf}
\usepackage{epstopdf}
\usepackage{multirow}
\usepackage{color}
\definecolor{Red}{rgb}{1.0,0,0}
\definecolor{Blue}{rgb}{0,0,1.0}
\newcommand{\be}{\begin{equation}}
\newcommand{\ee}{\end{equation}}
\newcommand{\bea}{\begin{eqnarray}}
\newcommand{\eea}{\end{eqnarray}}

\begin{document}
\title{Cold heteromolecular dipolar collisions}
\author{Brian C. Sawyer}
\author{Benjamin K. Stuhl}
\author{Mark Yeo}
\affiliation{JILA, National Institute of Standards and Technology
and the University of Colorado \\ Department of Physics, University
of Colorado, Boulder, Colorado 80309-0440, USA}
\author{Timur V. Tscherbul}
\altaffiliation{ITAMP, Harvard-Smithsonian Center for Astrophysics, Cambridge, Massachusetts 02138, USA}
\affiliation{Department of Physics, Harvard University, Harvard-MIT Center for Ultracold Atoms, Cambridge, Massachusetts 02138, USA}
\author{Matthew T. Hummon}
\affiliation{JILA, National Institute of Standards and Technology
and the University of Colorado \\ Department of Physics, University
of Colorado, Boulder, Colorado 80309-0440, USA}
\author{Yong Xia}
\altaffiliation{Department of Physics, East China Normal University, Shanghai 200062, China}
\affiliation{JILA, National Institute of Standards and Technology
and the University of Colorado \\ Department of Physics, University
of Colorado, Boulder, Colorado 80309-0440, USA}
\author{Jacek K{\l}os}
\affiliation{Department of Chemistry and Biochemistry, University of Maryland, College Park, MD 20742-2021, USA}
\author{David Patterson}
\author{John M. Doyle}
\affiliation{Department of Physics, Harvard University, Harvard-MIT Center for Ultracold Atoms, Cambridge, Massachusetts 02138, USA}
\author{Jun Ye}
\affiliation{JILA, National Institute of Standards and Technology
and the University of Colorado \\ Department of Physics, University
of Colorado, Boulder, Colorado 80309-0440, USA}
\date{\today}

\begin{abstract}
We present the first experimental observation of cold collisions between two different species of neutral polar molecules, each prepared in a single internal quantum state. Combining for the first time the techniques of Stark deceleration, magnetic trapping, and cryogenic buffer gas cooling allows the enhancement of molecular interaction time by 10$^5$. This has enabled an absolute measurement of the total trap loss cross sections between OH and ND$_3$ at a mean collision energy of 3.6 cm$^{-1}$ (5 K). Due to the dipolar interaction, the total cross section increases upon application of an external polarizing electric field. Cross sections computed from \emph{ab initio} potential energy surfaces are in excellent agreement with the measured value at zero external electric field. The theory presented here represents the first such analysis of collisions between a $^2\Pi$ radical and a closed-shell polyatomic molecule.
\end{abstract}

\maketitle

\begin{figure}[t]
\resizebox{9.0cm}{!}{
\includegraphics{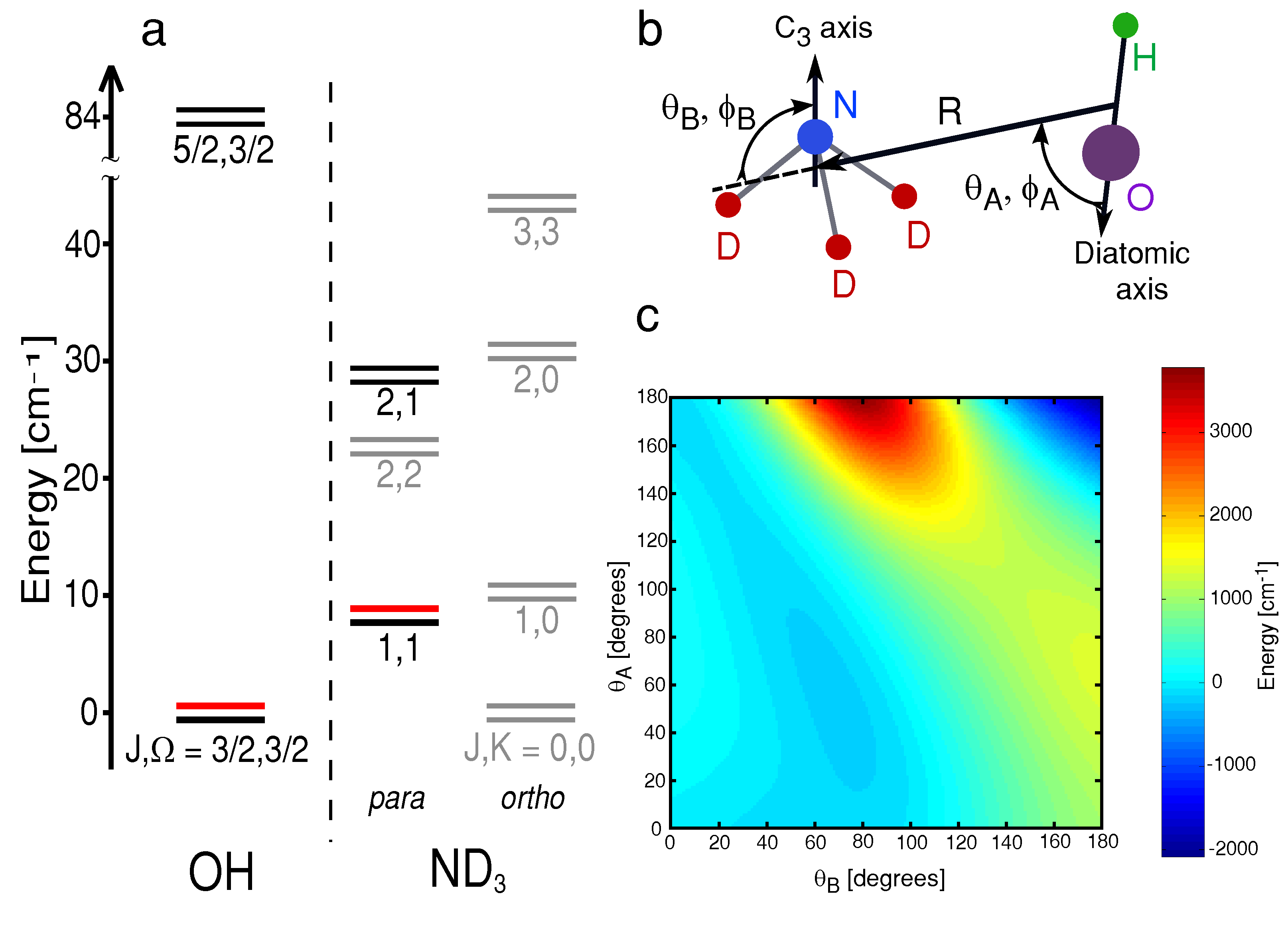}}
\caption{\label{fig1}\textbf{OH-ND$_3$ collision theory.} (a) Rotational structure of OH and ND$_3$ molecules. The parity-doublet splitting of each rotational level has been expanded for clarity, while the black and red parity levels are the ones included in the OH-ND$_3$ scattering calculation. The upper, red, parity states are those selected by the experimental apparatus for the cold collisions described herein. (b) Illustration of the Jacobi coordinates used for OH-ND$_3$ collision calculations. We define $R$ as the distance between the molecular centers of mass while the Euler angles ($\theta_A,\phi_A$) and ($\theta_B,\phi_B$) give the orientation of the OH and ND$_3$ axes, respectively, in the body-fixed frame relative to $R$. (c) Contour plot of the lowest adiabatic potential energy surface for the $A^{''}$ state of the OH-ND$_3$ collision complex at $R=3.1$ {\AA} and $\phi_A=\phi_B=0$ as a function of $\theta_A$ and $\theta_B$. The color legend is scaled in units of cm$^{-1}$.}
\end{figure}

For half a century, scattering of crossed atomic or molecular beams under single-collision conditions has remained the primary technique for investigation of inelastic and reactive dynamics at collision energies above $\sim$1 kcal/mol ($\sim$500 K)~\cite{Scoles88}. Experiments using electric field-aligned neutral polar molecules can probe the steric asymmetry of atom-molecule~\cite{vanBeek01,Alexander00,deLange04} and bi-molecular~\cite{Cireasa05,Taatjes07} potential energy surfaces (PESs), but their large center-of-mass collision energies have precluded the observation of electric field modification of these potentials. However, in the cold collision regime where only tens of scattering partial waves contribute, one can move beyond molecular orientation and directly control intermolecular dynamics via the long-range dipole interaction. The relatively low number densities (10$^6$-10$^8$ cm$^{-3})$ of cold molecule production techniques have thus far limited gas-phase collision experiments between state-selected distinct polar species to $>$300 K. We combine for the first time the methods of Stark deceleration, magnetic trapping, and buffer gas cooling to enhance molecular interaction time by $\sim$10$^5$ and overcome this density limitation. Here we present the first observation of cold (5 K) heteromolecular dipolar collisions and describe the first theoretical study of scattering in the many partial-wave regime with chemically-relevant OH and ND$_3$ molecules.

The last decade has seen tremendous progress in the production of cold and ultracold polar molecules~\cite{Carr09}. Application of these production techniques to the study of novel atom-molecule and molecule-molecule scattering has led gas-phase collision physics to new low temperature regimes. Specifically, crossed beam experiments using rare gas atoms and Stark decelerated OH ($^2\Pi$) molecules have allowed for the study of inelastic scattering at translational energies comparable to that of OH rotation ($\sim$100 K)~\cite{MeijerOHXe,Scharfenberg10,Kirste10}. By measuring collisional loss of a magnetically trapped OH target with incident beams of He or D$_2$, we have determined total collision cross sections at similar energies~\cite{Sawyer08}. Buffer gas cooling has been employed in the study of elastic and inelastic collisions of He-NH ($^3\Sigma$)~\cite{Campbell07} and He-TiO~\cite{Weinstein09}. Cold reactive collisions have been observed using a velocity-filtered room-temperature beam of CH$_3$F colliding with trapped Ca$^+$ atomic ions~\cite{Willitsch08}. In the ultracold regime, chemical reactions between ground-state KRb molecules at temperatures of a few hundred nanoKelvin have been directly controlled via external electric fields~\cite{Ospelkaus10,Ni10}, opening the door for experimental probes of quantum many-body effects in a dipolar molecular gas~\cite{Micheli06}.

\begin{figure}[t]
\resizebox{9.0cm}{!}{
\includegraphics{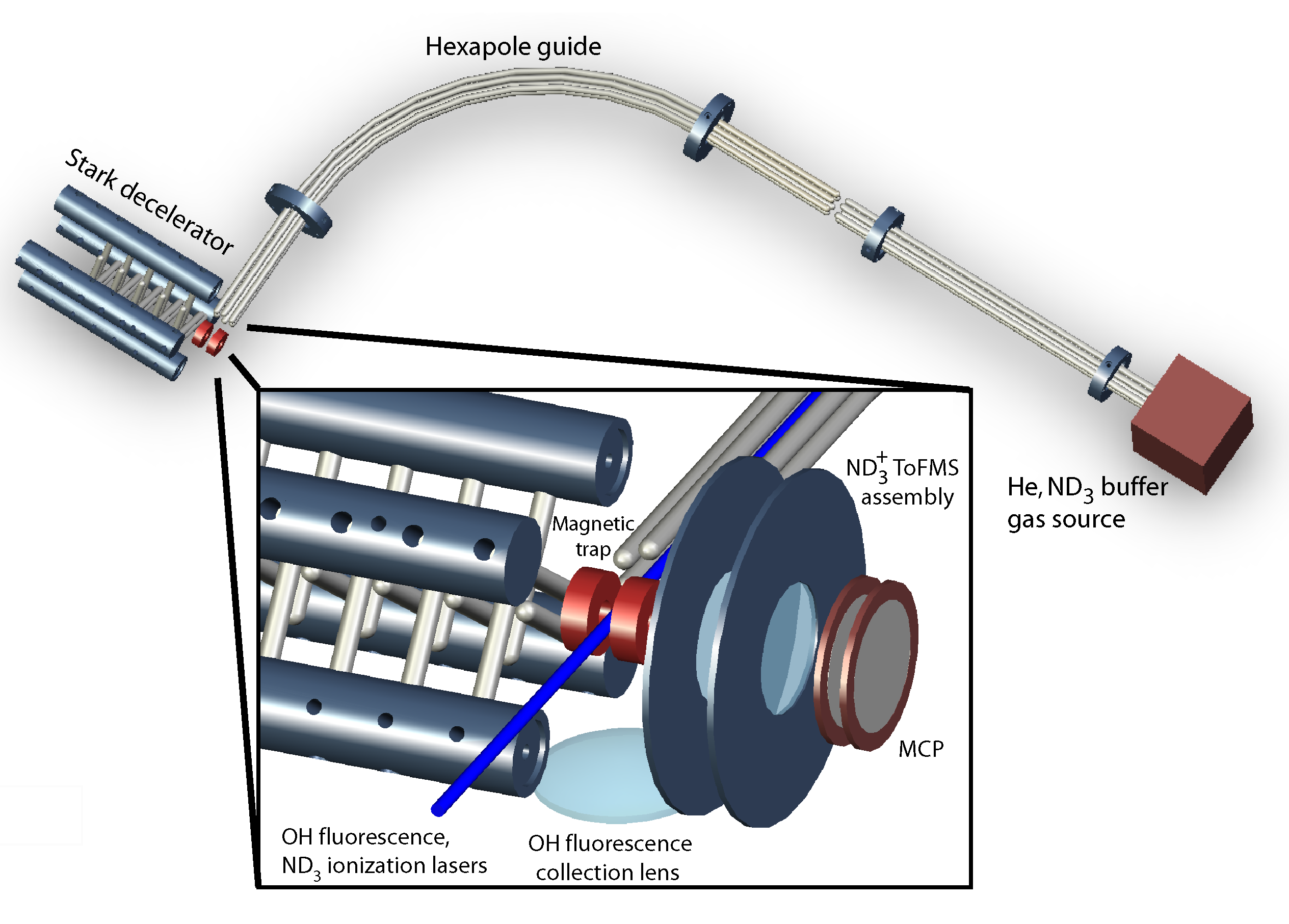}}
\caption{\label{machine}\textbf{Cold collision apparatus.} Illustration of the combined Stark decelerator, magnetic trap, and buffer gas beam assembly. The curved hexapole filters cold ND$_3$ from the He buffer gas and guides the continuous beam to the OH magnetic trap. (Inset) Closeup of the trap assembly showing the dual-species detection components. We detect OH and ND$_3$ in the collision region using laser-induced fluorescence (LIF) and resonance-enhanced multiphoton ionization (REMPI), respectively. Hydroxyl fluorescence at 313 nm is collected using a lens mounted 2.5 cm below the magnetic trap center. Ionized ND$_3^+$ molecules are accelerated to a microchannel plate (MCP) detector by placing 950 V on the front magnet, 0 V on the back magnet, and -1100 V on the acceleration plates that make up a time-of-flight mass spectrometer (ToFMS).}
\end{figure}

To observe dipolar collisions between state-selected OH and ND$_3$, we constructed a novel cold collision apparatus combining Stark decelerated and magnetically trapped OH with a continuous buffer gas cooled beam of ND$_3$, allowing interaction times of $\sim$1 s. In contrast to traditional scattering experiments, our measurement is sensitive to both inelastic and \emph{elastic} collisions. Furthermore, fewer than 100 partial waves contribute to the 5 K collision --- placing the OH-ND$_3$ interaction in an intermediate regime between gas-kinetic and quantum scattering. We report theoretical OH-ND$_3$ cross sections computed from \emph{ab initio} potentials that are in excellent agreement with the measured zero-field trap loss cross section. Given the generality of buffer gas cooling and electro/magnetostatic velocity filtering, our apparatus allows for a large class of electric field dependent cold molecular collision studies.

Two readily coolable species, hydroxyl and ammonia, are ubiquitous in atmospheric and astrophysical spectroscopy. Large interstellar clouds of OH and NH$_3$ (some co-located) have been detected in both absorption and maser emission along microwave parity-doublets~\cite{Lang80}. In the troposphere, hydroxyl-ammonia reactions are the dominant mechanism for removal of NH$_3$~\cite{McConnell73}. We choose to investigate electric field-dependent cold OH-ND$_3$ collisions due to the large static polarizability of both species and the near-degeneracy of the energy splittings of the OH $\Lambda$-doublet (0.056 cm$^{-1}$) and ND$_3$ inversion-doublet (0.053 cm$^{-1}$). The former allows for field orientation and precise control over the internal and external degrees of freedom of both molecules, while the latter has been shown to cause large enhancements in inelastic rates at room temperature~\cite{Ridley09}. Experimentally-relevant rotational and parity-doublet states of OH and ND$_3$ are illustrated in Fig.~\ref{fig1}a.

\begin{figure}[t]
\resizebox{6.0cm}{!}{
\includegraphics{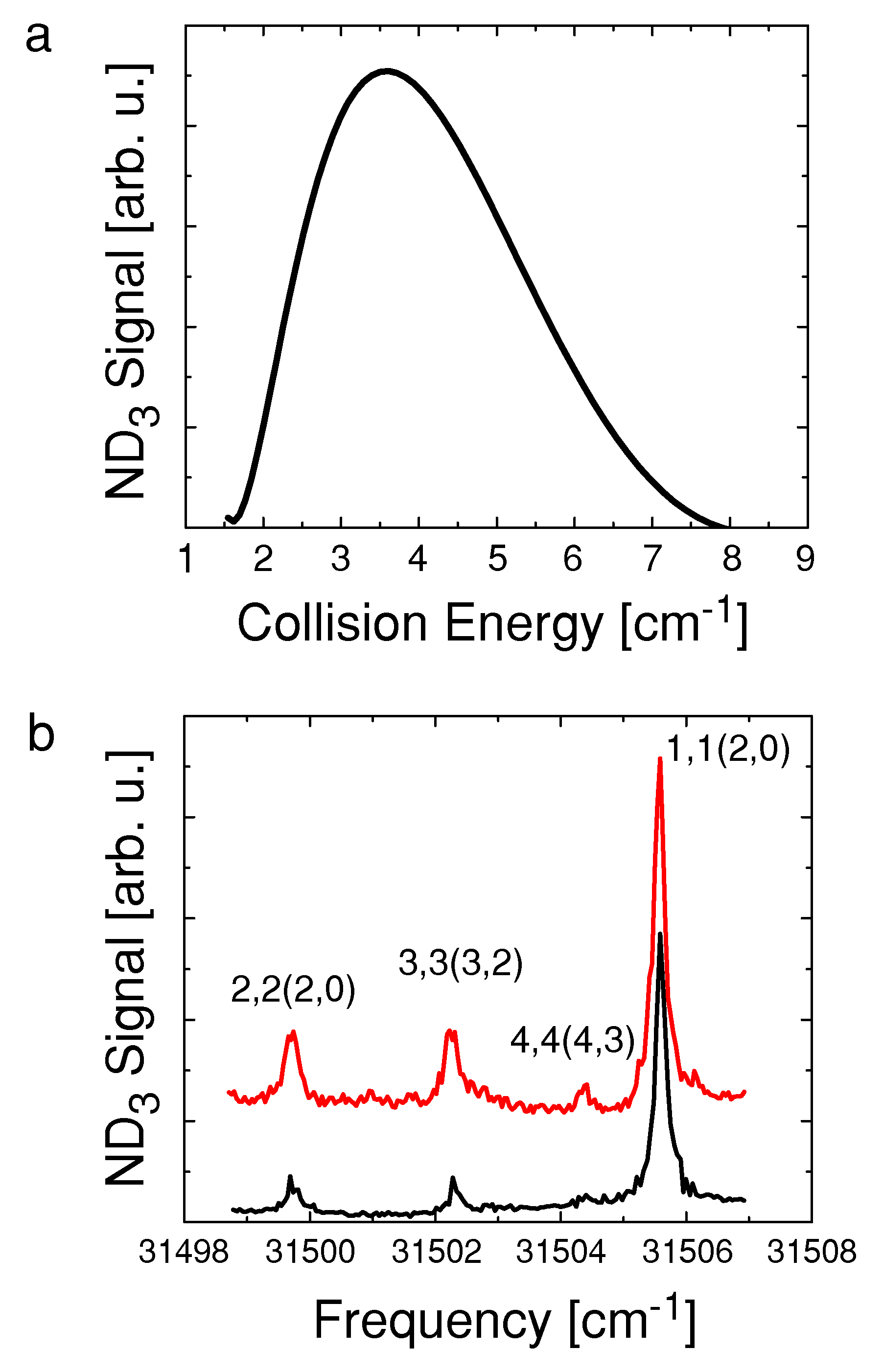}}
\caption{\label{spectra}\textbf{Translational and rotational ND$_3$ spectra.} (a) Translational energy spectrum of the guided continuous ND$_3$ beam as measured by 2+1 REMPI in the collision region. (b) Rotationally-resolved REMPI spectrum of guided ND$_3$ molecules showing different $J,K$ states. The upper red curve (offset for clarity) was taken at buffer gas flows of 2.0 and 2.5 sccm of He and ND$_3$, respectively. The lower black curve displays the smaller rotational temperature observed at 3.5 and 1.0 sccm. Intermediate rotational levels in the excited \~{B}($v_{2}=5$) electronic state are labeled in parentheses for each transition.}
\end{figure}

\textbf{Cold OH-ND$_3$ collision experiment}

The experimental setup consists of two distinct cold molecule apparatus as shown in Fig.~\ref{machine}: a Stark decelerator/magnetic trap assembly~\cite{SawyerMET07,Sawyer08} and a buffer-gas cooled continuous molecular beam source. A supersonic beam of OH molecules seeded in Kr is decelerated and trapped at a temperature of 70 mK and density of $\sim$10$^6$ cm$^{-3}$ as described in Ref.~\cite{Sawyer08}. Only those OH molecules residing in the $^2\Pi_{3/2}$ $|F,M_F,p\rangle=|2,+2,f\rangle$ ground state are phase-stably Stark decelerated and magnetically trapped. Here $F$, $M_F$, and $p$ denote the hyperfine quantum number, its projection on the external electric field axis, and the parity of the state, respectively. To detect trapped OH, the molecules are excited at the 282 nm $A^2\Sigma^{+}(v=1)$$\leftarrow$$X^2\Pi_{3/2}(v=0)$ transition. Fluorescence from the 313 nm $A^2\Sigma^{+}(v=1)$$\rightarrow$$X^2\Pi_{3/2}(v=1)$ decay is collected by a lens mounted to the magnetic trap assembly.

We employ buffer gas cooling and Stark velocity filtering to generate a continuous beam of cold ND$_3$ molecules at a density of 10$^{8}$ cm$^{-3}$ and mean velocity of 100 m/s. Our measured beam flux of 10$^{11}$ s$^{-1}$ is comparable to that of previous experiments~\cite{vanBuuren09,Patterson09}. Cold ND$_3$ is generated in a copper cell filled with He buffer gas at 4.5 K. A 6 mm-diameter aperture is cut into the front of the cell directly opposite the ND$_3$ inlet to allow the cold He/ND$_3$ mixture to escape. An electrostatic hexapole consisting of six 3 mm-diameter steel rods and possessing an inner diameter of 6 mm is mounted 2 mm from the cell aperture. This straight hexapole guide is 20 cm long and mounts to a 3 mm-thick gate valve which isolates the OH trap vacuum from the cryogenic vacuum. A curved hexapole is mounted just beyond the gate valve and follows a 90$^\circ$ bend at a radius of 13.5 cm. The hexapole guide terminates 1 cm from the permanent magnetic trap center. The curved hexapole filters cold ND$_3$ from the He buffer gas and sets an upper bound on the accepted forward velocity for weak-field seeking states ($\sim$150 m/s at $\pm$5 kV for $|J,K\rangle=|1,1\rangle$). The quantum numbers $J$ and $K$ denote the total molecular angular momentum and its projection on the ND$_3$ symmetry axis, respectively. Figure~\ref{spectra}a displays the measured energy distribution of guided $|1,1\rangle$ ND$_3$ molecules in the OH-ND$_3$ center-of-mass frame. To quantify the continuous beam velocity, the hexapole is switched so that guided ND$_3$ flux can be measured over different guiding durations. When the guide path length is included, the resulting curve is the \textit{integral} of the ND$_3$ velocity distribution. Guided ND$_3$ is detected in the collision region using 2+1 resonance-enhanced multiphoton ionization (REMPI)~\cite{Bentley00} and subsequent ion detection. To accomplish this, ND$_3$ in the $|1,1\rangle$ state is resonantly ionized between the trap magnets using a focused 317.4 nm, 10 ns laser pulse. To accelerate ions to a microchannel plate (MCP) detector (see Fig.~\ref{machine}), the trap magnets are charged to a potential difference of 950 V. After extraction from the trap, ions enter a 2 cm long field-free region and are subsequently detected by the MCP.

Our dual-detection scheme permits characterization of the state purity of both OH and ND$_3$. Due to the state selectivity of Stark deceleration and magnetic trapping, perfect OH rotational, $\Lambda$-doublet, and hyperfine state purity is achieved. Vibrationally excited ($v=1$) OH are phase stably decelerated with $v=0$ molecules, but $<$5\% of OH are initially produced in the $v=1$ state~\cite{Bochinski04}. We also characterize the ND$_3$ rotational distribution within the collision region. The rotational structure of ND$_3$ is shown in Fig.~\ref{fig1}a while Fig.~\ref{spectra}b displays the measured rotational spectra of guided ND$_3$ molecules for two different buffer gas flows -- 2.0 sccm He and 2.5 sccm ND$_3$ (upper curve) and 3.5 sccm He and 1.0 sccm ND$_3$ (lower curve). As the ratio of He to ND$_3$ flow is increased, the two-state ($|1,1\rangle, |2,2\rangle$) rotational temperature drops from 6.3 K to 5.4 K. With the larger He:ND$_3$ flow ratio, the relative guided populations of anti-symmetric (weak-field seeking) $|1,1\rangle$, $|2,2\rangle$, $|3,3\rangle$, and $|4,4\rangle$ states are 87\%, 9\%, 4\%, and $<$1\%, respectively. This ground state fraction is similar to that measured by depletion spectroscopy of a buffer gas cooled H$_2$CO beam~\cite{vanBuuren09}. States with $K=0$ do not exhibit a first-order Stark shift and are not guided. Also, at this flow and guide voltage ($\pm$5 kV), no population is observed in those states where $J$$\neq$$K$. Our choice of 3.5/1.0 sccm for the collision experiment is a compromise between maximizing $|1,1\rangle$ flux while minimizing rotational temperature.

\begin{figure}[t]
\resizebox{6.0cm}{!}{
\includegraphics{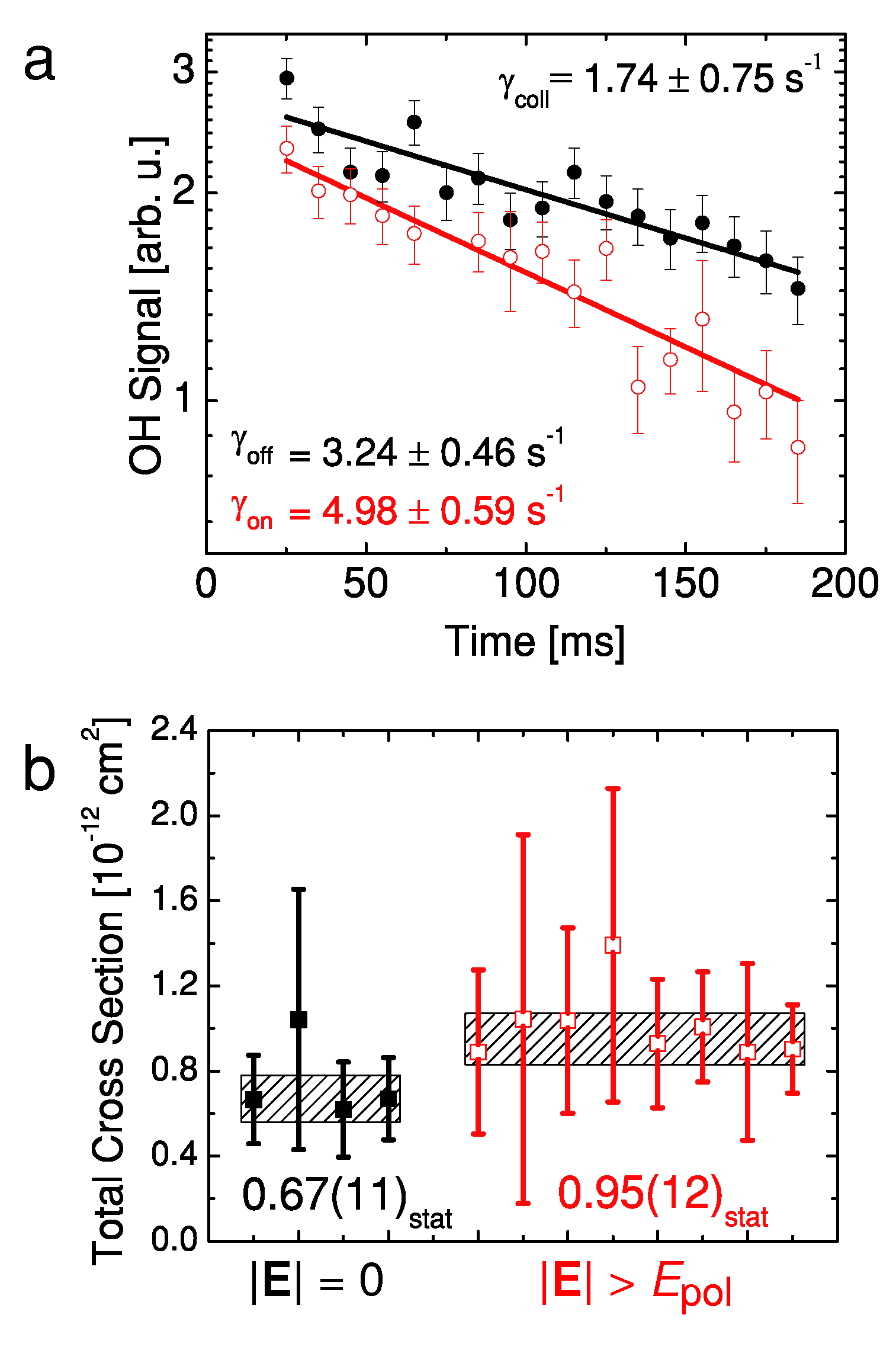}}
\caption{\label{sig}\textbf{OH collisional trap loss measurement.} (a) Semi-logarithmic plot of OH trap decay rates with ({\color{red}{$\circ$}}) and without ($\bullet$) the colliding ND$_3$ beam. The decay rate due solely to cold OH-ND$_3$ collisions is $\gamma_{\text{coll}}=\gamma_{\text{on}}-\gamma_{\text{off}}$. (b) Plot of all experimental runs measuring total cross sections with ({\color{red}{$\square$}}) and without ($\blacksquare$) a polarizing electric field. Average cross sections are determined from the weighted mean of all points and errors for the given $E$-field condition. The cross-hatched regions represent one statistical standard error.}
\end{figure}

To quantify total collision cross sections between trapped OH and guided ND$_3$, it is necessary to measure both collision-induced OH trap loss and absolute colliding ND$_3$ density. Loss of trapped OH is measured as shown in Fig.~\ref{sig}a over a period of 160 ms. First, the background-gas limited trap decay rate ($\gamma_{\text{off}}$) is measured with the ND$_3$ flow off. This decay rate is determined by a single-exponential fit to the data. The value of $\gamma_{\text{off}}$ is set by the presence of residual Kr in the trap chamber and is typically 2-3 s$^{-1}$ at a pulsed-valve repetition rate of 5 Hz. The cold ND$_3$ beam is then turned on and the enhanced loss rate ($\gamma_{\text{on}}$) is measured. The collision-induced loss rate ($\gamma_{\text{coll}}$) is the difference of the two rates. This differential measurement makes $\gamma_{\text{coll}}$ insensitive to day-to-day variations in background pressure, and yields a single-measurement fractional error of $\sim$30-40\% in the available runtime of 1 hr. To investigate the effect of an external electric field on the cross section, we apply two distinct electric field configurations to the colliding molecules. For both field distributions, $E>E_{\text{pol}}=\Delta/2\mu$ throughout the collision region, where $\mu$ is the permanent dipole moment and $\Delta$ is the parity-doublet splitting ($E_{\text{pol}}$=1.7 kV/cm for OH and 1.95 kV/cm for ND$_3$). One $E$-field distribution lowers the OH trap depth by $\sim$20\%, while the other increases the depth by $\sim$40\%. We observe no difference in collisional loss between these two configurations, and thus the changes in trap loss arise directly from the enhanced dipolar interactions between OH-ND$_3$ under the applied $E$-field.

%To investigate the effect of an external electric field on the cross section, we apply 1250 V across the magnet surfaces 1 ms after the OH is loaded into the trap. With this applied voltage, the average (minimum) electric field magnitude, $|\overline{\textbf{E}}|$, within the collision region of 2.6 kV/cm (2.0 kV/cm) is sufficient to polarize both OH and ND$_3$ throughout. The inhomogeneous external $E$-field causes a 20\% drop in OH trap depth, but this increases the 5 K collisional loss by a mere $\sim$1\%. Therefore, the observed $E$-field dependence arises purely from the modified intermolecular potential.

To determine the value of the loss cross section, $\sigma_{\text{exp}}^{\text{loss}}$, we use REMPI to calibrate the absolute density of the continuous ND$_3$ beam within the collision region for each measurement of $\gamma_{\text{coll}}$.  Conversion of MCP output current to absolute molecule density is notoriously difficult due to unknown ionization volume/efficiency and uncertain MCP gain. To sidestep these issues, a leak valve and calibrated quadrupole mass spectrometer are used to admit a known pressure of 295 K ND$_3$ gas into the trap chamber. The measured ND$_3$ pressure is then scaled down by the Boltzmann fraction of molecules in the $|1,1\rangle$ anti-symmetric state at thermal equilibrium (4.7x10$^{-3}$), and an absolute ND$_3$ density is assigned to the observed MCP output current. Provided the cold beam is larger than the ionization volume (true in our system), this calibration procedure is insensitive to the above problems. We also correct for overlap of REMPI transitions in the 295 K ND$_3$ spectrum. The relatively small rotational energy splittings of ND$_3$ displayed in Fig.~\ref{fig1}a lead to a congested spectrum at room temperature. Failure to account for overlap of transitions through the intermediate \~{B}($v_{2}=5$) state would lead to an underestimate of the cold beam density and concomitant overestimate of cross sections. Finally, a total trap loss cross section is determined for each experimental run as $\sigma_{\text{exp}}^{\text{loss}}=\gamma_{\text{coll}}(n_{0}\text{v}_{\text{rel}})^{-1}$, where $n_0$ is the measured density of ND$_3$ and $\text{v}_{\text{rel}}$ is the mean relative molecule velocity.

\textbf{\emph{Ab initio} calculations and quantum theory of OH-ND$_3$ collisions}

A complete theoretical description of the OH-ND$_3$ collision at 5 K is complicated by the $^2\Pi$ symmetry of ground-state hydroxyl. Interactions including $\Pi$ molecules are characterized by multiple electronic potentials, avoided crossings, and conical intersections. As such, scattering calculations based on \emph{ab initio} potential energy surfaces have thus far been limited to $\Sigma$ molecules~\cite{Cybulski05,Rios09} or collisions between $\Pi$ molecules and rare gas atoms or closed-shell diatomics with simple rotational structure. A full description of the theory results summarized here will be published elsewhere~\cite{TBP}.

To describe the OH-ND$_3$ collision complex, we calculate the two lowest adiabatic PESs of $A'$ and $A''$ symmetries using a spin-Restricted Coupled Cluster method with Single, Double, and non-iterative Triple excitations (RCCSD(T)) as implemented in MOLPRO~\cite{MOLPRO}. An augmented correlation-consistent triple zeta basis (aug-cc-pVTZ) is used for both N and O atoms~\cite{Dunning89}. As illustrated in Fig.~\ref{fig1}b, we define $R$ as the distance between the molecular centers of mass while the Euler angles ($\theta_A,\phi_A$) and ($\theta_B,\phi_B$) give the orientation of the OH and ND$_3$ axes, respectively, in the body-fixed frame relative to $R$. The interaction energies are evaluated using 7 Gauss-Legendre quadrature points in $\theta_A$ and $\theta_B$ and 13 equally spaced points in the dihedral angle $\phi=\phi_A-\phi_B$. To reduce computational cost of the {\it ab initio} calculations, we assume that the interaction potential is invariant under the internal rotation of the ND$_3$ monomer within the complex. This approximation is justified since no other states of \emph{p}-ND$_3$ are energetically accessible in the experimentally relevant range of collision energies (see Fig.~\ref{fig1}a), and \emph{para-ortho} interconversion is heavily suppressed in gas-phase collisions. Figure~\ref{fig1}c shows the lowest adiabatic PES of $A''$ symmetry as a function of the orientation angles $\theta_A$ and $\theta_B$ for $R=3.1$ \AA~and $\phi=0$.  Following the diagonal line defined by $\theta_A=\theta_B$, a barrier of $\sim$1000 cm$^{-1}$ is observed between head-to-tail ND$_3$$\cdot\cdot\cdot$OH and D$_3$N$\cdot\cdot\cdot$HO configurations, demonstrating the large anisotropy of the OH-ND$_3$ interaction.

To elucidate the dynamics of cold OH-ND$_3$ collisions, we developed and implemented a rigorous quantum scattering approach based on the close-coupling (CC) expansion of the wave function of the collision complex using the total angular momentum representation in the body-fixed coordinate frame. The matrix elements of the Hamiltonian are evaluated analytically by expanding the angular dependence of the {\it ab initio} PES in spherical harmonics and retaining only the lowest-order terms, including the isotropic term $V_{000}(R)$ and the dipolar term $V_{112}(R)$. The model PES produced in this way correctly reproduces both the isotropic and long-range anisotropic parts of the OH-ND$_3$ interaction, which are of crucial importance for cold collisions. Test calculations show that short-range anisotropic terms $V_{110}(R)$ and $V_{101}(R)$ have a minor influence on collision dynamics above 1 K.

The CC equations are solved numerically using the improved log-derivative algorithm to produce converged $S$-matrix elements and scattering cross sections for collision energies between 0.5 and 8 cm$^{-1}$, which are convoluted with the beam distribution function shown in Fig.~\ref{spectra}a to enable direct comparison with experimental data. The cross sections for trap loss are evaluated by integrating the differential cross section (DCS) over a restricted angular range $[\theta_\text{min},\pi]$. The cutoff angle $\theta_\text{min}$ serves to subtract contributions from forward-scattered elastic collision products that do not have enough kinetic energy to leave the trap.

\begin{figure}[t]
\resizebox{6.5cm}{!}{
\includegraphics{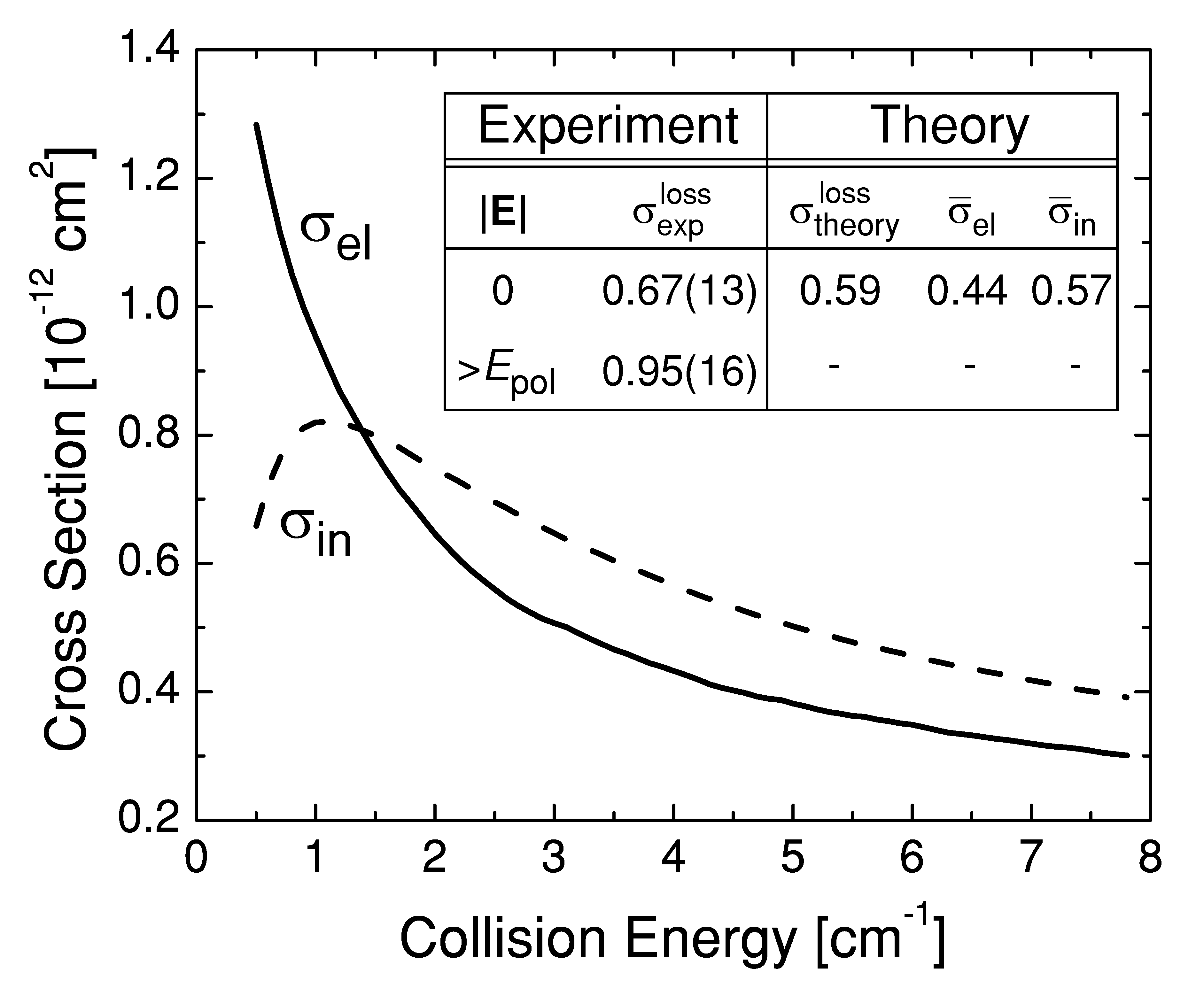}}
\caption{\label{SigmaTheory}\textbf{Theoretical and experimental OH-ND$_3$ cross sections.} Plot of theoretical OH-ND$_3$ elastic ($\sigma_{\text{el}}$) and inelastic ($\sigma_{\text{in}}$) cross sections over the experimental collision energy range. (Inset) Theoretical and experimental trap loss cross sections, with numbers in parentheses representing combined statistical and systematic errors. Experimental loss cross sections $\sigma_{\text{exp}}^{\text{loss}}$ are shown for both unpolarized and polarized colliding molecules. The $\sigma_{\text{theory}}^{\text{loss}}$ value includes the effect of reduced elastic loss from glancing collisions due to trap confinement. All cross sections are given in units of 10$^{-12}$ cm$^2$. The quantities $\bar{\sigma}_{\text{el}}$ and $\bar{\sigma}_{\text{in}}$ are the theoretical elastic and inelastic cross sections, respectively, in the absence of trapping potentials and averaged over the experimental collision energy distribution of Fig.~\ref{spectra}a.}
\end{figure}

\textbf{Results and Discussion}

The measured total trap loss cross sections are displayed in Fig.~\ref{sig}b. Each point represents a distinct cross section measurement and the average results for zero and non-zero electric field are determined as the weighted mean of each data set. The cross-hatched areas illustrate one statistical standard error centered at the weighted mean of each set. In addition to the statistical error, we estimate that the ND$_3$ leak pressure calibration and 295 K REMPI line-overlap correction add a common 11\% systematic uncertainty to both absolute measurements.

Figure~\ref{SigmaTheory} shows the calculated cross sections for elastic scattering and inelastic relaxation in OH-ND$_3$ collisions as functions of collision energy ($\mathcal{E}_C$). The cross sections are extremely large, exceeding typical gas-kinetic values by orders of magnitude. This dramatic enhancement is a direct manifestation of the dipolar interaction, which induces direct couplings between the opposite parity levels. Because of the long-range couplings, both elastic and inelastic cross sections are dominated by collisions occurring at large impact parameters. A partial-wave analysis of the cross sections at $\mathcal{E}_C=5$ K reveals contributions from as many as 90 partial waves, and the dependences $\sigma_{\text{el}}(J_{tot})$ and $\sigma_{\text{in}}(J_{tot})$ peak at $J_{tot}=30.5$ and $J_{tot}=50.5$, respectively, where $J_{tot}$ is the total angular momentum of the collision complex. At $\mathcal{E}_C>1$ K, the cross sections decrease monotonically with increasing $\mathcal{E}_C$, following the dependence $\sigma(\mathcal{E}_C)\propto\mathcal{E}_C^{-2/3}$. This result can be obtained analytically using the Langevin capture model \cite{BellSoftley} for a purely dipolar potential, demonstrating that the variation of the cross sections with collision energy is determined by the dipolar interaction.

A more detailed theoretical analysis shows that the inelastic cross section associated with both OH and ND$_3$ changing their parity-doublet state is $0.34\times10^{-12}$ cm$^2$, more than 6 times larger than that for only one partner changing state. This propensity rule follows from the symmetry properties of the matrix elements of the dipolar interaction in the scattering basis \cite{TBP}. The inset table of Fig.~\ref{SigmaTheory} compares the measured and theoretical trap loss cross sections. The $\sigma_{\text{theory}}^{\text{loss}}$ number is calculated using the OH-ND$_3$ differential cross section (DCS) and includes the recently-elucidated effect of trap confinement on measured collisional loss at low momentum transfer~\cite{Sawyer08,Fagnan09,TrapLoss}. Since glancing elastic collisions can leave scattered OH trapped, the measured $\sigma_{\text{exp}}^{\text{loss}}$ must be compared with $\sigma_{\text{theory}}^{\text{loss}}=\bar{\sigma}_{\text{el}}'+\bar{\sigma}_{\text{in}}$, where $\bar{\sigma}_{\text{in}}$ is the velocity-averaged inelastic cross section and $\bar{\sigma}_{\text{el}}'$ is the velocity-averaged elastic loss cross section that includes the trap suppression effect ($\bar{\sigma}_{\text{el}}'/\bar{\sigma}_{\text{el}}=0.04$). The forward-peaked structure of the OH-ND$_3$ DCS is responsible for this large reduction in elastic loss. Our measurement validates the large theoretical prediction of $\bar{\sigma}_{\text{el}}+\bar{\sigma}_{\text{in}}=1.0\times10^{-12}$ cm$^2$ attributed to the OH-ND$_3$ dipolar interaction.

We observe enhancement of $\sigma_{\text{exp}}^{\text{loss}}$ by a factor of 1.4(3) when both molecules are polarized by an external $E$-field. This signals an increase of the total cross section ($\bar{\sigma}_{\text{el}}+\bar{\sigma}_{\text{in}}$) since experimentally we observe no dependence of loss on trap depth. More work is required to compute a field-dependent $\sigma_{\text{theory}}^{\text{loss}}$ from the \emph{ab initio} PESs. However, we can estimate field-dependent elastic cross sections using the semiclassical Eikonal approximation as described in Methods. The result of this calculation combined with the above values of $\bar{\sigma}_{\text{in}}$ and $\bar{\sigma}_{\text{el}}'/\bar{\sigma}_{\text{el}}$ predicts a factor of 1.2 enhancement of the measured loss cross section when both molecules are polarized. This suggests that the elastic cross section plays a dominant role in the measured increase in trap loss.

\textbf{Conclusion and Outlook}

In summary, we present the first measurement of cold collisions between two distinct polar molecules in a variable electric field. Observing the total cross section increase under the polarizing $E$-field, we demonstrate preliminary control over the OH-ND$_3$ interaction at 5 K. Theoretical cross sections generated from \emph{ab initio} PESs agree well with the measured zero-field value and a semiclassical approximation reproduces the loss cross section for the case of polarized molecules. Future experiments will include detection of inelastic collision product states and OH microwave state transfer to a magnetically-confined lower $\Lambda$-doublet level for measurement of parity-dependent collision rates. Differential cross sections may also be directly mapped via systematic reduction of the OH trap depth with larger external electric fields. A variety of OH collision partners can be generated by the buffer gas source, allowing for a host of future cold collision studies between chemically-interesting molecules.

\textbf{Methods}

\textbf{Semiclassical estimate of field-dependent elastic OH-ND$_3$ cross section.} The Eikonal approximation assumes that scattering particles maintain a straight-line path throughout the collision process. This approximation is valid when the collision energy is larger than the scattering potential~\cite{Sakurai}. Assuming an isotropic and attractive Van der Waals potential of the form $-C_6/R^6$ without the applied $E$-field, where $C_6=21,380$ a.u. is computed from the dynamic polarizability of OH and ND$_3$, we obtain $\bar{\sigma}_{\text{Eik}}=0.50\times10^{-12}$ cm$^2$. In the presence of the $E$-field, we assume a dominant dipole-dipole potential of the form $\langle\mu_{OH}\rangle\langle\mu_{ND3}\rangle[1-3\cos^2\theta]/R^3$, where $\theta$ is the angle between the molecules' symmetry axes and $\langle\mu_{OH}\rangle=1.0$ D, $\langle\mu_{ND3}\rangle=0.77$ D are the expectation values of the OH and ND$_3$ permanent electric dipoles, respectively~\cite{Bohn09}. This interaction gives $\bar{\sigma}_{\text{Eik}}^{\text{dip}}=3.0\times10^{-12}$ cm$^2$. Including the values of $\bar{\sigma}_{\text{in}}=0.57\times10^{-12}$ cm$^2$ and $(\bar{\sigma}_{\text{el}}'/\bar{\sigma}_{\text{el}})=0.04$ obtained from the \emph{ab initio} PESs, semiclassical theory predicts a factor of 1.2 enhancement in the measured loss cross section when both species are polarized --- in agreement with the measured value. It is possible that $\bar{\sigma}_{\text{in}}$ will not change in the presence of the external $E$-field since the dipole potential contributes weakly at short-range.

\textbf{Buffer gas source.} The copper buffer gas cell is mounted to the second stage of a closed cycle pulse-tube refrigerator. Under typical operating conditions, the temperatures of the second stage and cell are 4.5 K. Pre-cooled He gas flows into the side of the cell through 2.4 mm diameter copper tubing. The warm (280 K) ND$_3$ tube is mounted to the back of the cell and is thermally isolated via a thin-walled 2.5 cm diameter tube of epoxy/fiberglass composite with a mating polyetherimide insert. The ND$_3$ tube reaches a base temperature of 240 K, and 360 mW is applied to a resistive heater to achieve a 280 K operating temperature. Given that ammonia freezes below 195 K, ice formation within the buffer gas cell is unavoidable. However, we achieve stable beam operation over $>$1 hour by maximizing the distance ($>$1 cm) between the warm ND$_3$ inlet and cold cell walls. When ND$_3$ ice finally bridges the gap between the cell wall and warm inlet, sudden vaporization of the ice causes a large pressure rise in the cell and surrounding chamber which leads to significant cell heating and loss of cold beam flux. Charcoal sorb ($\sim$2000 cm$^2$) at 4.5 K is used to pump He gas within the dewar vacuum.

\textbf{Acknowledgments}

We acknowledge DOE, AFOSR-MURI, NSF, and NIST for funding support. M. Hummon is a National Research Council postdoctoral fellow. T.V. Tscherbul was supported by NSF grants to the Harvard-MIT CUA and ITAMP at Harvard University and Smithsonian Astrophysical Observatory. We thank G. Qu\'{e}m\'{e}ner and J. L. Bohn for stimulating discussions.

%\bibliography{MagTrap}

\end{document}